# Metal Domain Size Dependent Electrical Transport in Pt-CdSe Hybrid Nanoparticle Monolayers

*Michaela Meyns,[§] Svenja Willing, Hauke Lehmann and Christian Klinke**

Institute of Physical Chemistry, University of Hamburg,

Grindelallee 117, 20146 Hamburg, Germany

[§]Present address: Catalonia Institute for Energy Research (IREC), Jardins de les Dones de Negre 1, 08930 Sant Adrià de Besòs, Barcelona, Spain

*email: klinke@chemie.uni-hamburg.de

*Abstract*

Thin films prepared of semiconductor nanoparticles are promising for low-cost electronic applications such as transistors and solar cells. One hurdle for their breakthrough is their notoriously low conductivity. To address this, we precisely decorate CdSe nanoparticles with platinum domains of one to three nanometers in diameter by a facile and robust seeded growth method. We demonstrate the transition from semiconductor to metal dominated conduction in monolayered films. By adjusting the platinum content in such solution-processable hybrid, oligomeric nanoparticles the dark currents through deposited arrays become tunable while maintaining electronic confinement and photoconductivity. Comprehensive electrical measurements allow determining the reigning charge transport mechanisms.

*Keywords*: Colloidal hybrid nanoparticles, electrical transport, photoconductivity, size effects, Langmuir-Blodgett deposition

Continuous progress in the control of nanoparticle (NP) size, composition, and shape has led to unprecedented possibilities in material design targeting applications as diverse as medical therapy and diagnostics and solid state devices.[1] Nanoparticle thin-film devices are studied and applied in various contexts, especially with regard to their optoelectronic and electronic properties.[2] The solution processability of nanoparticle building blocks makes them particularly attractive for applications such as photovoltaics,[3-6] light emitting diodes,[7-9] as well as chemical sensing and photodetection.[10-14] The fundamental electrical properties of single material systems consisting of either semiconductor or metal nanoparticles have been studied especially with respect to quantum-mechanical coupling phenomena,[15-20] Coulomb-blockade behavior,[21,22] transistor characteristics,[23-25] and photo conductivity.[20,26,27]

Nanoparticles prepared by colloidal chemistry can be arranged into ordered arrays by a variety of methods. Among them are the widely applied spin-coating, self-assembly,[28,29] and the Langmuir-Blodgett technique.[30] In particular, the latter relies on a stabilizing layer of organic surfactants that enables the dispersion and assembly of nanoparticles on a liquid sub-phase. While highly ordered films can be achieved, a big drawback of this method is the creation of spatial and energetic barriers between adjacent nanoparticles by the organic stabilizers. This usually leads to a high resistivity of the arrays.[31] Several approaches have been developed to improve the conductivity of nanoparticle thin films. The most popular ones are heat treatment and ligand exchange of already deposited arrays.[2,(32-35] Recently, methods have been reported where ligands are introduced prior to deposition that decompose upon annealing of the array and leave the surface basically uncapped.[36,37]

Another option to influence the conductivity of semiconducting nanoparticles is the carefully adjusted addition of another material. This may happen *via* engineering of the composition, by doping or the formation of heterostructures.[19,38-42] Colloidal oligomeric hybrid nanoparticles (HNPs) offer the possibility to tailor material combinations for advanced applications in the fields of photocatalysis and (opto-)electronics.[43,44] A number of studies have demonstrated the excellent photo-catalytic properties of metal-semiconductor HNPs in liquid solution, which are based on an efficient separation of photogenerated charge carriers.[45] The electrons move towards the metallic part of the HNPs from where they can be transferred to reducible species in the solution. Pt-semiconductor combinations belong to the most powerful photocatalysts.[44]

Despite the plentitude of photocatalytic studies and the often applied description of HNPs as promising candidates for optoelectronic applications, only few studies have investigated their electrical characteristics in solid-state devices in detail. Depending on the combination of materials, the characteristics of HNP thin films vary significantly. In thin films of Au-PbS core-shell structures, for example, electrons are transferred to the metallic core, thus supporting hole conductivity.[46] FePt cores with Cd or Pb chalcogenide shells, however, did not exhibit such behavior. Instead, a tunneling process of charge carriers between the metallic cores was proposed for transport through these NPs.[47] In single Au-tipped CdSe nanorods an increase in the conductance of five to six orders of magnitude compared to bare nanorods was observed.[41] CdSe nanorod arrays and networks decorated with and/or welded by Au spheres exhibited enhanced dark currents and photocurrents that are mediated by the morphology of the hybrid structures.[42,48-50] Heterostructures of $Cu_2ZnSnS_4$ (CZTS) and Au were recently reported to improve the photoresponse and -stability compared to pure CZTS multi-layered photodetectors.[51]

Increasing the current in a semiconductor nanoparticle array by deposition of metallic domains seems intuitive. Anyhow, it is less clear how the metal domain size and thus the degree of coupling between the domains influences the dark and especially the photo-currents. In the following, we report on the synthesis of well-defined oligomeric Pt-CdSe HNPs by a simple seeded-growth approach and the effect of Pt domain size on their electrical transport characteristics in darkness and under illumination in monolayer devices.

**Results and discussion**

*Oligomeric Pt-CdSe hybrid nanoparticles*

Oligomeric Pt-CdSe HNPs with several Pt domains grown onto a larger CdSe nanocrystal were prepared by a straightforward and robust seeded-growth approach in organic solution. To examine the nature and tuneability of electrical transport in two-dimensional arrays of Pt-CdSe HNPs, the size of the metal domains was varied systematically. Quasi-spherical chalcogenide nanoparticles with crystallographically exposed sites proved to be excellent materials for metal deposition.[52,53] For this reason, hexagonal pyramidal CdSe NPs were synthesized as seeds.[54] Pt(acac)$_2$ was dissolved in toluene and oleylamine was added to act both as ligand and reducing agent. CdSe NPs dispersed in toluene were added at room temperature before the mixture was heated to reflux. To modify the Pt domain size, the reaction time was varied while the ratio of Pt atoms to CdSe NPs in solution was maintained (Pt atoms : CdSe NPs = 39000 : 1). **Figure 1** shows the products of such syntheses. The combination of the Pt precursor, oleylamine, and nanoparticles is not reactive towards hybrid formation at room temperature (see Supporting Information). Even under toluene reflux there is only a slow growth rate of the metallic domains. As indicated by the high-resolution images in Figure 1b and c, the long reaction time allows for a controlled crystallization of Pt and the formation of well-defined interfaces. In contrast to what has been observed in case of Au deposition, we observed no signs for ripening of metal domains with prolonged reaction times.[55] *In situ* heating of the NPs during electron microscopy confirmed the stability of the heterostructures up to temperatures of 300 °C. Further selected-area electron diffraction, XRD data and EDX mapping underlining the crystallinity and distinguishability of the two components can be found in the Supporting Information.

With respect to the optical properties of the HNPs shown in **Figure 2**, increasing the Pt-domain size leads to less defined CdSe UV/Vis absorbance bands as well as quenching of the photoluminescence. The maxima of the first optical transition and the emission shifted insignificantly. The observed changes are typical for metal-semiconductor nanomaterials with interfacial contact.[44,55-58] They are attributed to electronic interactions between the two domains or even charge transfer from the semiconductor to the metal domain by non-radiative processes and additional light absorption of the metallic component.[59-62] In solution based studies, HNPs with Pt domains were shown to more efficiently separate and transfer photogenerated charge carriers compared to hybrids with other metals.[58,63-66] The advantage of Pt-CdSe and Pt-ZnO HNPs compared to those with Au is a reduced charging of the metal domains in combination with an efficient transfer of electrons towards redox active species in the surrounding reaction medium.[58,66] In CdSe, for example, a faster charge transfer from semiconductor to Pt than to Au caused by an altered interface barrier is in accordance with properties of macroscopic contacts between these materials: a lower Schottky barrier has been observed with the first material combination.[67] Fast transfer of the electrons that have been photogenerated in the semiconductor towards the metallic domain can be expected to reduce the recombination probability of the charge carriers and thus, be beneficial for light-charge transducer devices like photodetectors and solar cells.

*Hybrid nanoparticle monolayer device preparation*

For the electrical investigations CdSe NPs with average dimensions of 10.7 nm in diameter and 12.3 nm in the *c*-axis were decorated with Pt domains of 1.0 nm, 1.9 nm, and 2.8 nm in diameter during a seeded-growth process with reaction times of 4 h, 14 h, and 24 h respectively. The corresponding EDX data shown in **Table 1** documents the variation in Pt content from below 1% (1.0 nm) to 21% (2.8 nm). When Pt is deposited, a slight loss of Cd occurs, a circumstance that has also been observed upon metal deposition onto other Cd-chalcogenides.[68] This is considered as related to the partial reduction of the metal precursor by the chalcogenide compound that is necessary to form seeds for the growth of metallic domains and a concomitant release of surface Cd atoms. Nevertheless, the decrease in semiconductor diameter, typically occurring during redox-deposition of metals determined by TEM was within the experimental and methodological error. In the following, for simplicity we will use rounded values for the Pt domain size: 1, 2, and 3 nm. HNPs with Pt domains of these sizes were deliberately chosen since the first is in the cluster regime where energy levels are quantized,[69] the second is in the transition region between clusters and metallic domains (around 2 nm), and the last is well in the metallic regime. The purified NPs were assembled and deposited onto Au electrode structures. The 25 nm high electrodes were pre-patterned on n-doped Si substrates isolated by a 300 nm thick thermal oxide layer. The films were prepared on electrodes with interdigitating fingers of 25 μm in length separated by a gap of 450 nm (**Figure 3a**). Electrodes and contact pads were defined by electron-beam lithography, thermal evaporation of 2 nm Ti as adhesion layer and 23 nm Au and a final lift-off step. Each of the four synthesized particle types, pure CdSe NPs and the different Pt-decorated ones, were assembled to monolayers by the Langmuir-Blodgett technique using diethylene glycol as sub-phase[30] and finally deposited on top of the Au electrodes as monolayers as shown in Figure 3a and b (for TEM images see Supporting Information).

*Dark currents as a function of metal content*

We investigated the electrical transport properties of homogeneous monolayers of pure and Pt-decorated CdSe NPs respectively under high-vacuum conditions of about $10^{-5}$ mbar. The advantage of examining monolayers is that only two-dimensional transport has to be taken into account when analyzing transport properties. This simplifies the analysis of processes in a film of nanostructures with a morphology and composition that is intrinsically complex already. An important question to be answered in the following is the nature and strength of the contribution of Pt to the transport. As indicated in Figure 3d, two scenarios are possible: In one case, charge carriers move across interfaces under participation of CdSe energy bands, in the other one they only move between Pt domains *via* tunneling.

At 300 K, the current-voltage characteristics (I-V curves) show expectedly high resistance in the examined bias voltage region for pure CdSe-NP films measured in darkness (**Figure 4**, for non-logarithmic plots see Supporting Information). The applied bias voltage ranged from -20 V to +20 V, corresponding to fields of up to approximately $40 \cdot 10^6$ V/m between the electrodes. In films of HNP with Pt clusters of 1 nm (<1% Pt in the HNP; about 35 Pt atoms) the I-V curves in the dark are similar to those obtained with pure CdSe films. With 2 nm Pt domains (6 % Pt content in total) there is a significant dark current in the pA range at an applied bias voltage of ± 10 V, while currents of tens of nA are observed with 3 nm Pt domains (21 % Pt content in total) for the same applied bias voltage. A small hysteresis, visible as an incongruence of the I-V curves during cycling the bias voltage from -20 V to +20 V and back to -20 V, was observed for low Pt contents. The reason for its occurrence may be charging effects during the course of one specific I-V cycle and is typically recorded in semiconductor NP arrays.[70] With the biggest Pt domains on the other hand, such a hysteresis is not resolved, indicating a more metal-like behavior of the array. This suggests that surface traps and other defect sites in the semiconductor itself do not play an important role anymore, due to the already high current through the film.

A comparison of the electrical (Figure 4) and the optical characterization (Figure 2) reveals that those HNPs with larger Pt domains conduct above noise level at room temperature that show a total quenching of the photoluminescence. Vice versa, if a distinct luminescence is observed like in particles with small or no Pt domains no current through the particles can be measured in the applied bias-voltage regime. This can be understood by assuming a dominating role of Pt domains in the transport process. Large Pt domains lead to a more metallic character which promotes the electrical transport. If the Pt domains are small and cluster-like, their capacitance to accommodate or transfer charge carriers is low such that the semiconducting component plays a bigger role in the overall HNP properties and quenching is incomplete due to a lower contribution of alternative pathways.

To further investigate the influence of Pt domains and examine transport processes that may occur, we compared the dark current measurements at a temperature of 300 K with those carried out at 7.5 K. The samples with little or no Pt show a high resistance at both temperatures and no current is measured in the applied source-drain voltage interval between ± 20 V. In contrast, the currents recorded for samples with 2 nm and 3 nm Pt domains are about one order of magnitude lower for 7.5 K compared to room temperature. Further, the current is more affected by temperature in the regime between -10 V and +10 V than at higher biases.

The positive part of the I-V curves measured with HNPs containing 3 nm Pt is plotted and complemented by fit functions in Figure 4c and d. Above a bias voltage of ±10 V (E = $20 \cdot 10^6$ V/m) the current–voltage characteristics obtained at room temperature cannot be attributed to one specific process as fits for different mechanisms converge (Figure 4c). Anyhow, the comparatively small change of the current through the samples at high applied bias voltages with temperature indicates that field driven tunneling processes with negligible temperature dependence dominate the electrical transport in the HNPs.

At 300 K and voltages below 10 V, fitting for simple thermionic emission with a dependence of $\exp(B \cdot V^{1/2})$ (were B is a fitting constant) is not satisfactory ($\chi^2=1.8 \cdot 10^{-19}$), whereas the curve can be fitted with a $\exp(B \cdot V^{1/4})$ function ($\chi^2=9.3 \cdot 10^{-21}$). Such proportionality is characteristic for thermionic emission processes across Schottky barriers at macroscopic metal-semiconductor interfaces under an applied reverse voltage and the influence of an image charge.[41,67] This may seem surprising in so far as the length scale of the depletion layer that is present on the semiconductor side at macroscopic interfaces may equal or exceed the dimensions of the nanostructures. However, the same exponential dependence was found for single CdSe nanorods tipped with Au[41] and Pt contacted GaN nanowires around room temperature and at low biases.[71] Theoretical and STM studies support the existence of highly localized interfacial dipoles and nanoscale Schottky barriers in metal tipped nanowires of different materials longer than 1 nm.[72-75]

In a thermionic emission process across Schottky barriers under reverse bias the current exponentially depends on the height of the barrier at the interface and the thermal energy under consideration of the local electric field. The corresponding relations can be written as

$$I = AR^*T^2 exp\left(\frac{-q\Phi_{BE}}{k_BT}\right) \quad (1)$$

with an effective barrier height

$$\Phi_{BE} = \Phi_{B0} - \sqrt{\frac{qE}{4\pi\epsilon_S}} \quad (2)$$

and a maximum field [67,71]

$$E = \sqrt{\frac{2qN_D}{\varepsilon_S}\left(V + \Phi_{bi} - \frac{k_BT}{q}\right)} \quad (3)$$

In these equations $A$ is the electrically active contact area, $R^*$ the effective Richardson constant, $q$ the elementary charge, $T$ the temperature, $k_B$ the Boltzmann constant, $E$ the maximum electric field at the metal-semiconductor junction, $\varepsilon_S$ the semiconductor permittivity, and $N_D$ the doping concentration in the semiconductor. The built-in potential $\varphi_{bi}$ is determined by the energetic difference between the semiconductor conduction band at the interface and inside the crystal.

A voltage driven emission of thermally activated electrons from trapped states or NPs as described by Frenkel and Poole with currents proportional to V·exp(B·V$^{1/2}$) is also a possible mechanism for the superlinear increase of the current.[67] Though this proportionality fits a bit better ($\chi^2=1.5\cdot10^{-21}$) than the thermionic emission with mirror charges, it is less likely since trap emission requires high fields.[42]

At 7.5 K there is a conductance gap where the current signal falls under the noise level in the voltage regime of ± 2.5 V (Figure 4d) for 3 nm Pt domains and between ± 10 V for 2 nm Pt domains. We attribute this gap and its size dependence to a Coulomb blockade, in agreement with observations made with CoPt metal NPs.[21] The curve of HNPs with 3 nm Pt domains was analyzed and fitted to a $V^2\cdot\exp(-B/V)$ function, which gives evidence for tunneling as the dominating transport process at low temperatures.[67]

All the above observations let us propose a transport model for high contents of Pt where emission and tunneling processes play the major role. At low temperatures tunneling dominates the electrical transport through the HNPs. At room temperature and low bias voltage thermal emission of electrons from metallic to semiconducting domains governs the transport and CdSe plays a role in the paths of the electrons, as depicted in the upper part in Figure 3d. With increasing electric fields, transport along the metallic domains (or sub-gap states) and tunneling from particle to particle aided by the field induced reduction of inter-particle barriers become dominant (lower part in Figure 3d).

*Photocurrents*

To examine the influence of the metal content on the photoinduced current flowing through the film, laser irradiation with a wavelength of 637 nm and with an intensity density of up to 203 mW/cm$^2$ was applied. The chosen wavelength is sufficient to overcome the band gap of 668 nm (1.86 eV) in the CdSe NPs but does not excite plasmonic resonances in Pt which lie in the UV range.[76] Upon irradiation, photocurrents were recorded in all samples, as plotted in **Figure 5a**. Bias voltages of above ±10 V were necessary to sufficiently separate and transport photogenerated charges in pure CdSe and those with cluster-sized Pt domains so that currents could be measured.

For pure CdSe and with 1 nm Pt domains, the photocurrent remains comparatively low, even at elevated bias voltages. However, the current induced by the irradiation accounts for almost 100 % of the total current since the dark current was close to the noise level. The similarity of the two samples once again shows that the Pt domains are too small to affect the transport characteristics significantly.

While preserving the clear photoresponse of a semiconductor, the HNPs with 2 nm Pt domains show an increased current even close to zero bias. The same accounts for the case with 3 nm Pt domains, in which the benefit of a higher dark current is compromised by a slower, less defined response and an increase of the dark current with time, likely linked to accumulation of carriers in the array. This may be explained by an increased capacity of the Pt domains and the HNPs in general. For an application in photodetectors, for example, a stable dark current and a large amplification upon irradiation is desired. The amplification factors of the photocurrent gain $I_{illuminated}-I_{dark}$ and the relative photocurrent $A_F = (I_{illuminated}-I_{dark}) / I_{dark}$ measured at an applied bias voltage of +10 V under irradiation with 203 mW/cm$^2$ derived from the data shown in Figure 5 are given in **Table 2**.

The highest photocurrent gain $I_{illuminated}-I_{dark}$ was obtained with 3 nm Pt. The highest relative photocurrent was calculated with 1 nm Pt domains, although the noise level was taken for the dark current. Pt domains of 2 nm lead to the highest relative photocurrent with a stable dark current and quick response. A higher photocurrent gain with larger Pt domains can be explained by a more effective charge transfer to Pt.[77,78] The On/Off ratio of the current through the NPs shown in Figure 5b–e increases with increasing laser intensity for all four films. A linear dependence between photocurrent and intensity occurs when exciton dissociation is the rate limiting step, meaning that many excitons recombine while they are still associated (geminate recombination).[79] Our findings suggest that, in order to achieve a material with an optimum ratio between photo and dark current, the size of the Pt domain must be adjusted to be above cluster size but small enough to maintain a stable and distinct response.

**Scheme 1** depicts the energetic levels and charge carrier paths for the extreme cases of CdSe without Pt under illumination and Pt-CdSe with 3 nm Pt in the dark. CdSe (and Pt-CdSe with small Pt domains) conducts only upon irradiation of the array, as expected for such a semiconductor NP film. With 3 nm Pt domains the conduction wins in metallic character. At low voltages charge transfer from one domain to the other is assisted by thermionic emission. At higher voltages tunneling processes are dominant. With larger Pt domains, the photocurrent is merely a small addition to the already high dark current. Thus, intermediate Pt sizes such as 2 nm would lead to transport properties between the two extremes with noticeable dark current and clear photoresponse.

**Conclusion**

By varying the Pt domain size the electrical conductance of thin films of colloidally stable oligomeric Pt-CdSe HNPs can be modulated without the need to remove the protecting organic ligand layer. With cluster sized Pt domains of 1 nm, electrical transport through the HNPs is similar to what is observed with pure CdSe samples. Increasing the size of the Pt domains leads to a stronger contribution to the transport.

Thermionic- and field-emission processes increase the dark currents at room temperature while the transport at low temperatures is dominated by tunneling. Both, optical and electrical properties of the HNPs, are strongly related to the size of the Pt domains. Particles with large Pt domains show a total quenching of the photoluminescence and reveal a comparatively high conductivity while particles without Pt or only small Pt domains yield a clear luminescence signal but conduct only at around the noise level.

For efficient photodetectors it is important to minimize the noise in the output signal. A distinguished photoresponse at low applied voltages with linear dependence on the irradiation intensity combined with a defined dark current qualifies the presented Pt-CdSe HNP as building blocks for photodetectors. Our results demonstrate that it is possible to increase the conductivity of semiconducting materials by variation of the metallic domains in size while maintaining the semiconducting character represented by their photoresponse. By carefully choosing other hybrid material combinations and adjusting the domain sizes it may thus be feasible to cover broad ranges of photodetection. Furthermore, an optimized electrical transport in combination with the high number of interfaces in oligomeric HNPs that can act as scattering centers for phonons may be interesting in terms of thermoelectric applications.

**Materials and Methods**

**Materials**

Cadmium oxide (CdO; 99.99+%) was purchased from ChemPur. Tri-*n*-octylphosphane (TOP 90% vacuum distilled) and selenium shots (amorphous, 2-4 mm, 0.08-0.16 in, 99.999+%), both stored in a glovebox with nitrogen atmosphere, were obtained from Sigma-Aldrich. Tri-*n*-octylphosphane oxide (TOPO; >98%), 1,2-dichloroethane (DCE; p.A.), chlorobenzene (DCE; p.A.), acetone (p.A.), methanol (p.A.), ethanol (p.A.), 2-propanol (p.A.), trichloromethane (p.A.), *n*-hexane (p.A.), diethylene glycol (p.A.), and toluene (p.A.) were acquired from Merck. Acros is the producer of our oleylamine (OA; 80-90% C18 content, stored under nitrogen atmosphere). *n*-Octadecylphosphonic acid was bought from PCI (ODPA; ≥99%), Platinum(II) acetylacetonate (Pt(acac)$_2$; 98%) from abcr.

The chemicals were used without further purification if not indicated otherwise.

**Methods**

*Synthesis of CdSe nanocrystals*

CdSe nanocrystals with wurtzite crystal structure and hexagonal bipyramidal geometry were synthesized following a previously published method with minor modifications.[54] In a three necked 25 mL flask with a condenser, a septum and a thermocouple in a glass mantle, 25 mg (0.19 mmol) CdO, 0.14 g (0.42 mmol) *n*-octadecylphosponic acid (ODPA), and 3.0 g (7.8 mmol) tri-*n*-octylphosphine oxide (TOPO) were heated to 120 °C for 30 minutes. During this time, two switches from vacuum to nitrogen and back were carried out. Under nitrogen flow the temperature was raised to 270 °C for 60 minutes for the complexation of Cd after which an optically clear and colorless solution was obtained. Stirring was maintained at 336 rpm (MR Hei-Standard stirrer) throughout the reaction. At 80 °C to 77 °C, 10 µL (0.13 mmol) of dichloroethane (DCE) were injected and the temperature was raised again. At 250 °C, 0.42 mL (0.42 mmol Se) of a Se in TOP solution (1 M) was injected before reducing the temperature to 240 ± 2 °C for growth. The color of the solution gradually turned to brown *via* orange and red, indicating the formation of CdSe NPs. After 4 hours the reaction was quenched by cooling to 70 °C and injecting 3.5 mL of toluene. The resulting CdSe NPs were purified by three cycles of precipitation with methanol (toluene/methanol 1:0.5), centrifugation (1856g, 3 min), removal of the supernatant, and re-dispersion in toluene. The optical density of the resulting stock solution was determined by UV-vis spectroscopy.

*Synthesis of Pt-CdSe hybrid nanoparticles*

Platinum acetylacetonate (24 mg, 62 µmol) was dissolved in toluene to give a clear yellow solution. To this solution, 2.0 mL of oleylamine (C18 content 80-90 %, 5.0 mmol for 80 %) were injected. After 5 minutes of stirring, an appropriate amount of CdSe NPs in toluene was added to result in a reaction solution of 25 mL with an optical density of 0.1 regarding CdSe (first absorption maximum) corresponding to a NP concentration of 64 nmol/L (for details see Supporting Information). The mixture was heated to reflux and left to stir for varied periods of time (RH basic 2 IKAMAG stirrer at speed 3). With proceeding reaction time, the color of the solution began to turn into a darker brown indicating the growth of Pt domains on the CdSe NPs. When the reaction was run at room temperature, no change in color was observed. The reaction was quenched by removing the heating mantle after the desired reaction time. The hybrid nanoparticles were precipitated with methanol (1:1.5), centrifuged for 3 min at 11504g and re-dispersed in 4 mL of toluene. The supernatant remained slightly colored due to small amounts of Pt clusters that had formed during the reaction. Aliquots taken during the reaction were further purified by precipitation and re-dispersion with methanol (1:1), trichloromethane/acetone (1:1) and again toluene/methanol (1:1). Centrifugation was run for 3 minutes at 11504g. For characterization, the nanoparticles were re-dispersed in toluene.

*Nanoparticle purification for monolayer deposition*

For one pure CdSe NP film, an equivalent of 1 mL (~1/7) of the CdSe NP reaction solution was purified. The sample was separated into two centrifuge vials (2 mL), 0.125 mL of methanol were added and the sample was centrifuged (3 min at 1856g). After re-dispersion and precipitation with toluene and methanol (0.5/0.75 mL) the NPs were centrifuged (3 min at 1856g) and stored under nitrogen until the sub-phase for the Langmuir-Blodgett process was prepared. The NPs were re-dispersed in 1 mL of toluene, centrifuged (3 min at 11092g) and precipitated from the remaining supernatant with 1 mL of methanol and centrifugation (3 min at 1856g; rotation at higher values was necessary on occasion). The purified NPs were taken up in 50 to 200 µL of toluene to give a dark brown dispersion.

For further purification of Pt-CdSe HNPs, 3/4 of the as prepared NPs were first precipitated with methanol (1:1) and centrifuged at 11504g for 3 min. They were taken up in the original volume of *n*-hexane and carefully centrifuged so as not to precipitate the particles but the appearing drop-sized liquid phase (3 min at 4754g). The supernatant was then transferred and precipitated with ethanol (1:1, 3 min at 11504g). The nanoparticles were re-dispersed in trichloromethane (5/7 of the original volume) and precipitated with acetone (1:1) and centrifugation at 11504g for 3 min. Finally, the nanoparticles were re-dispersed in 50-100 µL of toluene. Toluene is the solvent best-suited for spreading the nanoparticles on the diethylene glycol sub phase for Langmuir-Blodgett deposition due to its comparatively slow evaporation and its compatibility with the Teflon trough. Switching the solvent/anti-solvent combinations during the purification process was necessary to remove residues from the reaction in order to obtain dispersions well spreadable and ordering on the sub phase.

*Electrode preparation*

The electrical transport properties of hybrid nanoparticle monolayers were characterized with interdigitated array electrode structures. The electrodes were prepared on n-doped (Sb) Si substrates with a thermal oxide layer of 300 nm by electron beam lithography. For this, polymethylmetacrylate (PMMA) was spin coated onto the substrates (waiting time 90 s, rotation: 1 min at 4000 U) from chlorobenzene. Patterning was carried out at a Zeiss Supra 55 SEM using the CAD software *Elphy Quantum* (Raith). The dimensions were set to 0.5 µm for the distance between the fingers and 25 µm for their length. Ti (2 nm) and Au (23 nm) were vapor deposited with a Pfeiffer Vacuum Classic 250 system. The lift-off of sections coated by PMMA was carried out in acetone.

*Monolayer deposition*

Monolayer deposition was carried out with a KSV Nima KN-2002 system and Nima WIN LB software with diethylene glycol (DEG) as a sub phase. The electrode substrates were bathed in ethanol, isopropanol and demineralised water prior to use. After drying, they were mounted onto the workshop made sample holder, as described in reference (30), inserted into the holder of the dipper, and dipped into the sub phase with an optimum deposition angle of 15° between DEG surface and substrate (105° between substrate and sample holder). Several drops of the priory purified NP dispersion were spread out onto the Langmuir-Blodgett sub phase with a microliter glass syringe. They covered roughly two thirds of the space between the barriers. Once the nanoparticles were spread out on the sub-phase and the solvent had evaporated, they were compressed to a dense monolayer at a constant rate of 2 mm/min with a target pressure of 9 to 10 N/cm$^2$ and left to relax for up to 2 hours until the change of area became negligible or the formation of double layers close to the barriers was observed. At a rate of 1 mm/min the substrates were lifted out of the sub phase. They were then transferred into a vacuum oven, dried at room temperature over night at <10$^{-2}$ mbar and finally transferred directly to the probe station or stored in a nitrogen flooded cabinet.

*Electrical transport measurements*

Transport measurements were conducted in a VFTTP4 probe station from Lake Shore Cryotronics at 10$^{-5}$ to 10$^{-6}$ mbar with a Keithley 4200-SCS semiconductor characterization system or an Agilent 4156C Precision Semiconductor Parameter Analyzer, respectively. For measurements under irradiation a 637 nm laser with adjustable intensity was used.

*UV-vis spectroscopy*

Absorbance measurements were carried out with a Perkin Elmer Lambda 25 two-beam spectrometer. Emission spectra were obtained with a Horiba Jobin Yvon Fluoromax-4 spectrophotometer at an irradiation wavelength of 480 nm. For all measurements quartz cuvettes with an optical path length of 10 mm were employed.

*Transmission Electron Microscopy (TEM), Energy Dispersive X-ray Spectroscopy (EDX)*

Standard transmission electron micrographs were obtained with a JEOL JEM 1011 microscope with a thermal emitter operated at an accelerating voltage of 100 kV. High resolution TEM micrographs and EDX data were obtained with a JEOL JEM 2200FS (UHR) equipped with a field emitter, CESCOR and CETCOR correctors and a Si(Li) JEOL JED-2300 energy dispersive X-ray detector at an accelerating voltage of 200 kV. Furthermore, this microscope was employed for in-situ heating experiments with Pt-CdSe nanoparticles employing a heatable stage. Purified samples (10 µL) were drop-casted onto copper grids covered with amorphous carbon films. Medium values of nanoparticle dimensions were determined from at least 210 counts with the software *Image J*.

*Powder X-ray diffration*

XRD measurements were carried out with a Philips X'Pert PRO MPD with Bragg Brentano geometry and a Cu(K) X-ray source emitting at 0.154 nm. The data was analyzed with the *X'Pert Highscore Plus* software.

**Acknowledgements**

We would like to thank D. Weinert and A. Kornowski for HR-TEM, SAED, and EDX mapping and Y. Cai for her assistance with Langmuir-Blodgett deposition. Discussions about hybrid and metal nanoparticle synthesis with N. G. Bastús are highly appreciated. The authors thank the German Research Foundation DFG for granting the projects KL 1453/5-1 and KL 1453/9-1 and the European Research Council for funding an ERC Starting Grant (Project: 2D-SYNETRA, Seventh Framework Program FP7).

**Supporting Information Available**

Further methodological details, digital images and micrographs, EDX mapping, SAED, and XRD data as well as IV-curves in linear scale are provided. This material is available free of charge *via* the Internet at http://pubs.acs.org.

**Figures**

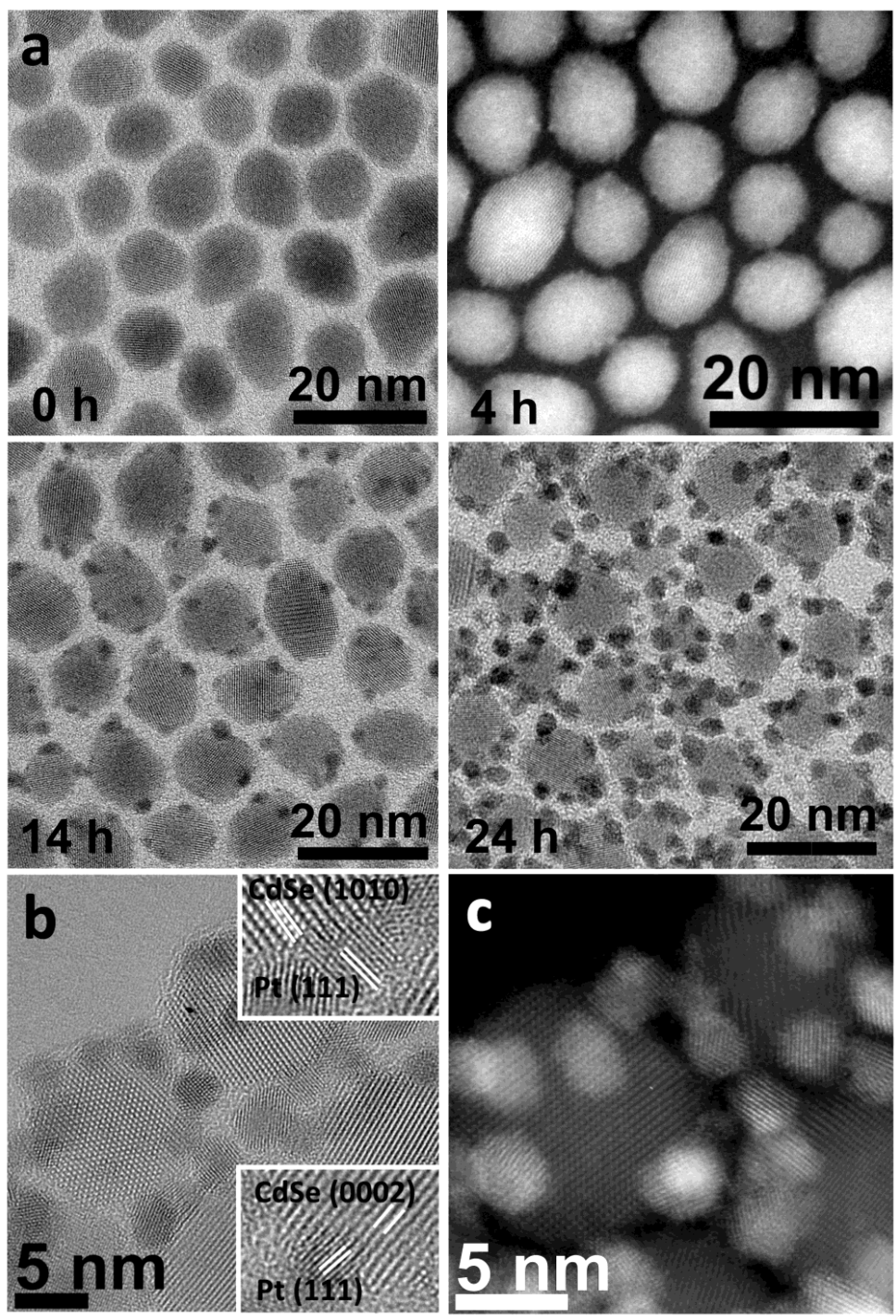

**Figure 1.** TEM and STEM micrographs of pure CdSe NPs and Pt-CdSe HNPs obtained at different reaction times. (a) The Pt-domain sizes are 1.0 ± 0.2 nm after 4 h, 1.9 ± 0.3 nm after 14 h, and 2.8 ± 0.5 nm after 24 h. The image for 1 nm Pt domains is taken in STEM mode since the clusters can barely be seen otherwise. (b) HR-TEM micrograph with lattice fringes of wurtzite CdSe and Pt. (c) HR-STEM image of Pt-CdSe NPs after 24 h of reaction demonstrating the crystallinity and well-defined morphology of the hybrid nanoparticles.

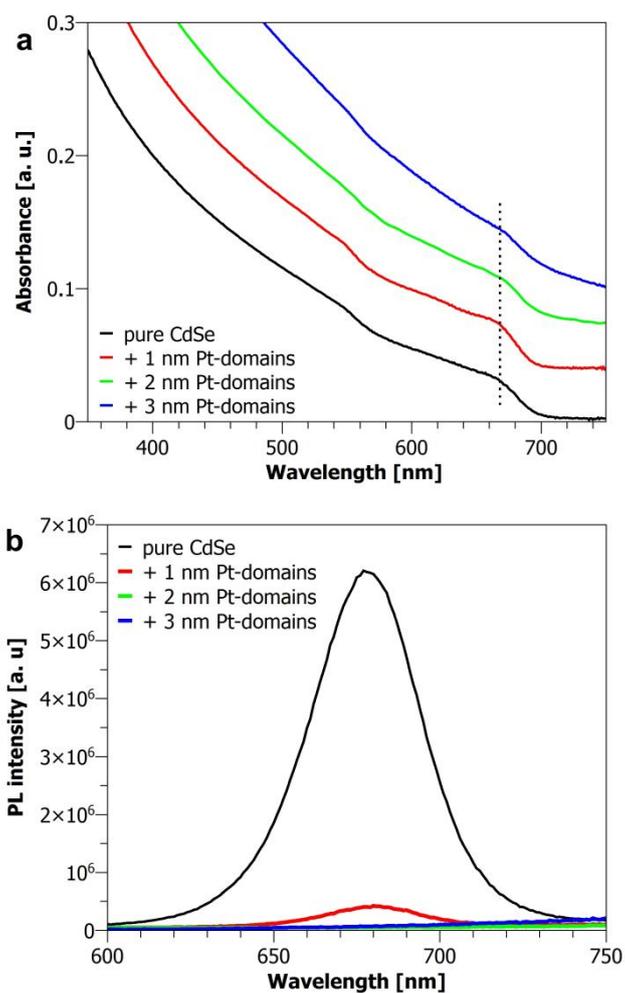

**Figure 2.** (a) Absorbance and (b) emission of pure and Pt-decorated CdSe NPs dispersed in toluene. The absorbance curves are offset for clarity. The absorbance maximum at 668 nm and the PL peak at 678 nm shift insignificantly upon metal deposition.

**Table 1.** Elemental composition and atomic Se/Cd ratios of pure and Pt-decorated CdSe NPs with different Pt diameters $D_{Pt}$ as determined by EDX.

| $D_{Pt}$ (nm) | EDX (atom %) | | | Atomic ratio Se/Cd |
|---|---|---|---|---|
| | Cd | Se | Pt | |
| - | 46 | 54 | - | 1.2 |
| 1.0 ± 0.2 | 42 | 58 | < 1 | 1.4 |
| 1.9 ± 0.3 | 39 | 55 | 6 | 1.4 |
| 2.8 ± 0.5 | 35 | 44 | 21 | 1.3 |

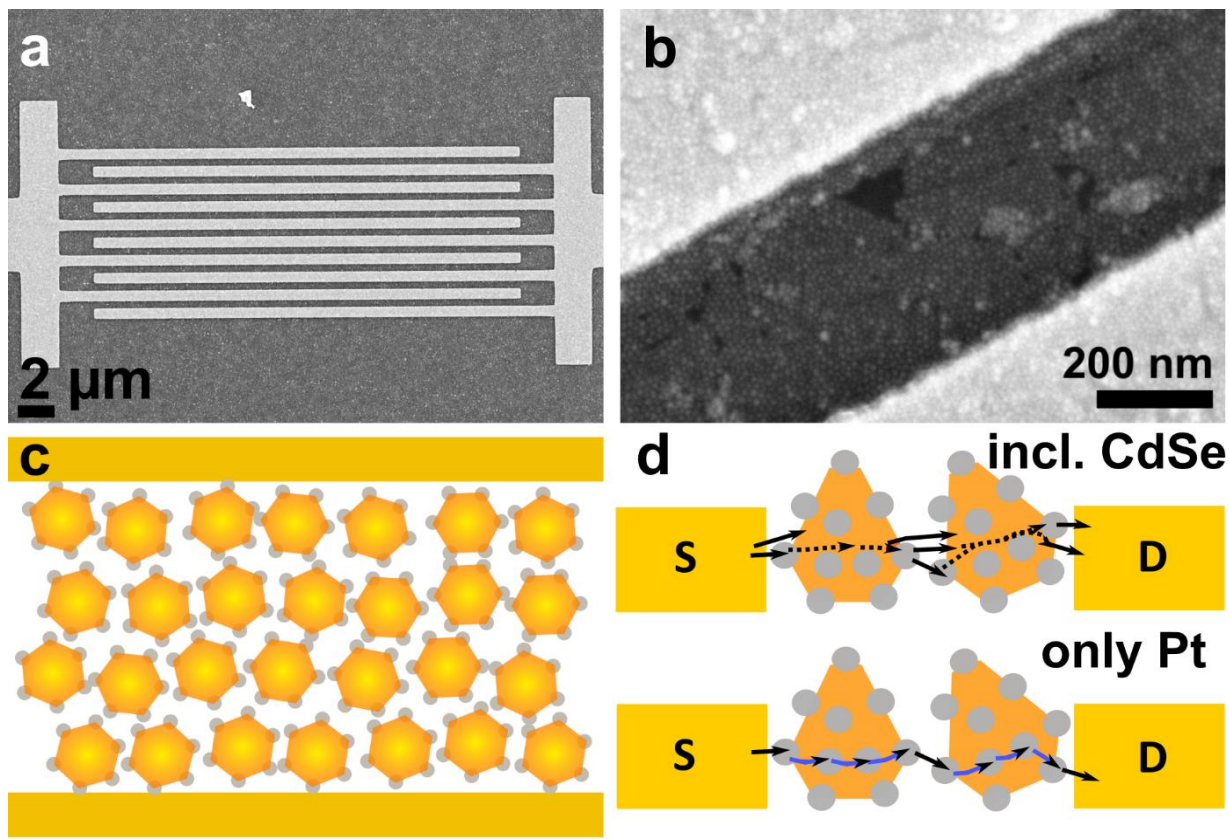

**Figure 3.** (a, b) SEM micrographs of interdigitated array electrodes with Pt-CdSe hybrid NPs and CdSe NPs. (c) Scheme of hybrid NPs between two electrode fingers and (d) possible paths for charge carriers from source (S) to drain (D) across the NP film: either including the semiconductor (top) or only passing *via* Pt (bottom).

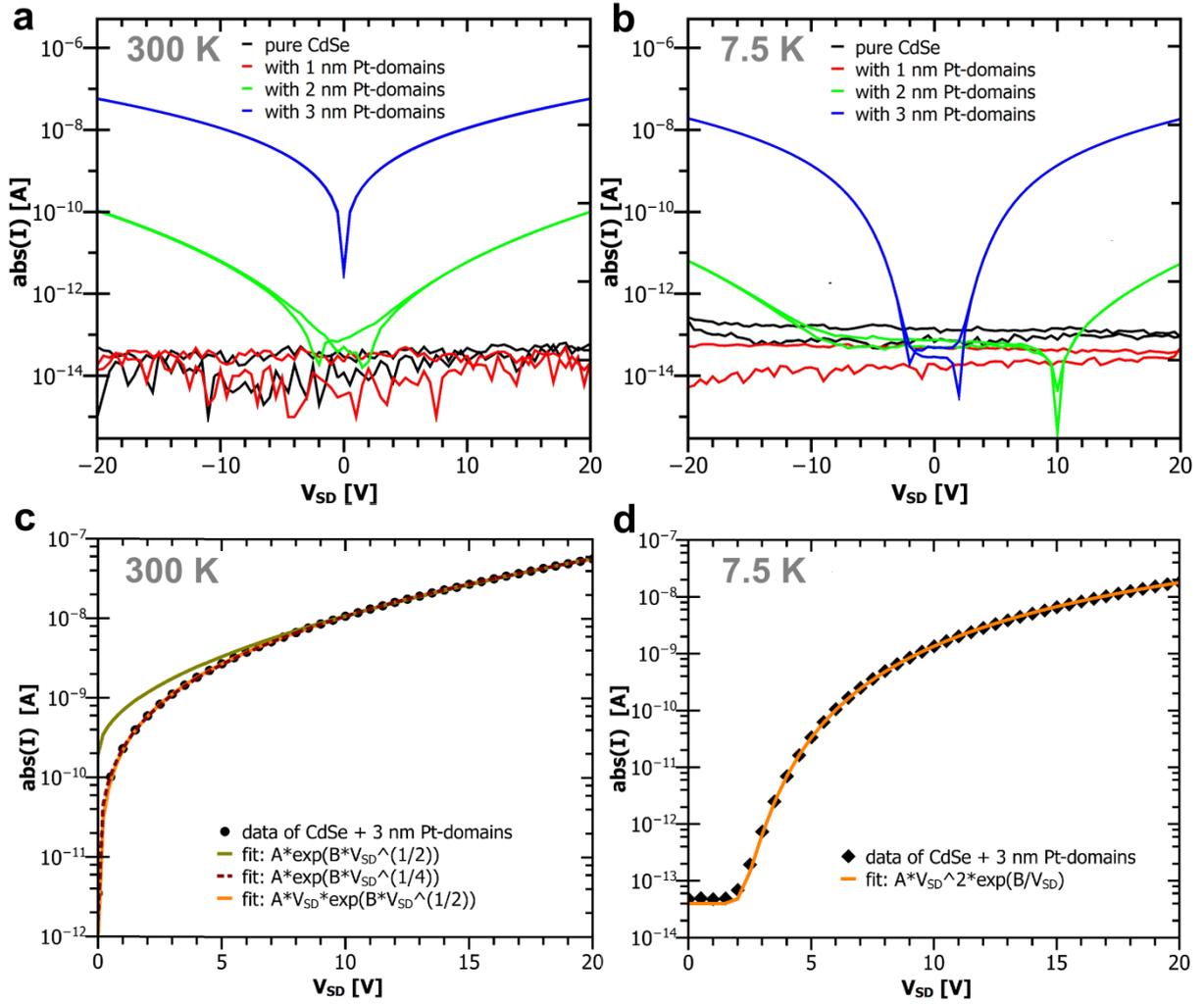

**Figure 4.** Dark currents through monolayers of pure and Pt-decorated CdSe at a temperature of (a, c) 300 K and (b, d) 7.5 K. The positive part of the current-voltage curves is fitted for CdSe NPs with 3 nm Pt domains. At (c) 300 K an exponential voltage dependence fits with thermionic emission processes whereas at (d) 7.5 K tunneling transport and a Coulomb gap occur. A and B represent fitting constants.

**Table 2.** Photocurrent gain and amplification factor for HNP arrays with different Pt domain diameters $D_{Pt}$ under irradiation with 203 mW/cm$^2$ at 637 nm and a bias voltage of +10 V.

| $D_{Pt}$ [nm] | Photocurrent gain ($I_{ill}$-$I_{dark}$) [pA] | Amplification factor $A_F$ |
|---|---|---|
| - | 0.135 | 24.9 |
| 1 | 0.267 | 33.4 |
| 2 | 5.00 | 0.667 |
| 3 | 100 | 0.009 |

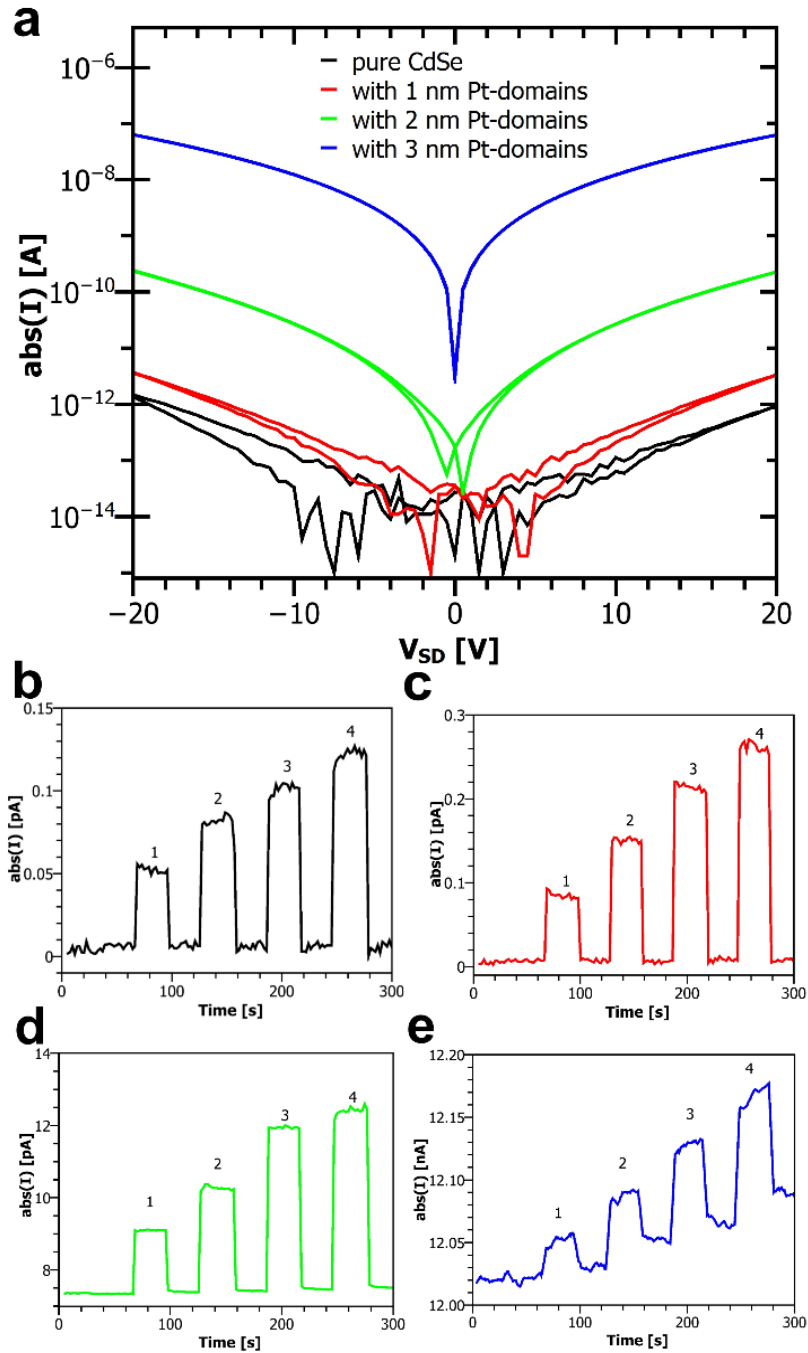

**Figure 5.** Room temperature measurements of the photocurrent through films of pure and Pt-decorated CdSe NPs under laser illumination (λ = 637 nm). (a) Absolute currents during laser illumination. Intensity dependency of the photocurrents in the films with (b) pure CdSe NPs, (c) 1 nm Pt-domains, (d) 2 nm Pt-domains, and (e) 3 nm Pt-domains. With increasing Pt domain size the underlying dark current increases and higher net photocurrents are obtained. A bias voltage of +10 V is applied to provide a separation of the charge carriers in all samples. Numbers denote the applied laser-intensity densities (1: 49 mW/cm$^2$, 2: 101 mW/cm$^2$, 3: 158 mW/cm$^2$, 4: 203 mW/cm$^2$).

**Scheme 1.** Simplified scheme of the proposed dominating charge transport processes, only taking into account the most likely paths. As extreme cases CdSe and Pt(3nm)-CdSe arrays are depicted. In CdSe and Pt(1nm)-CdSe current flows only upon irradiation. With large Pt domains transitions to CdSe across a barrier lower than that presented by vacuum or ligands occur at low bias voltages. At high bias voltages tunneling between Pt domains or sub gap states at the interface across barriers lowered by the applied field becomes more prominent. Photocurrents are just a small addition to the flowing dark currents. HNPs with 2 nm Pt range in between these two extremes.

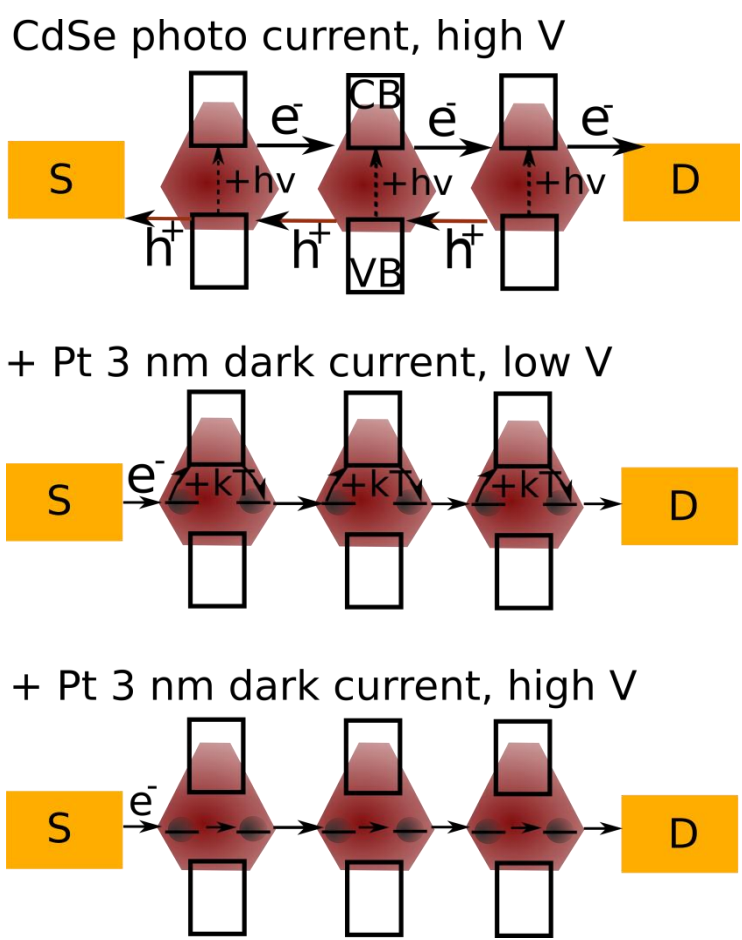